\long\def\symbolfootnote[#1]#2{\begingroup%
\def\thefootnote{\fnsymbol{footnote}}\footnote[#1]{#2}\endgroup} 
\def\aj{AJ}
\def\araa{ARA\&A}
\def\apj{ApJ}
\def\apjl{ApJ}
\def\apjs{ApJS}
\def\aap{A\&A}
\def\mnras{MNRAS}
\def\nat{Nature}

\newcommand{\dr}{{\rm d}}
\newcommand{\feh}{{\rm [Fe/H]}}
\newcommand{\moverl}{\Upsilon_V}
\newcommand{\moverlten}{\Upsilon_{V,10}}
\newcommand{\lv}{L_V}

\newcommand{\msun}{{\rm M}_{\odot}}
\newcommand{\mobs}{M_{\rm obs}}
\newcommand{\mssp}{M_{\rm SSP}}
\newcommand{\mlo}{m_{\rm lo}}
\newcommand{\mup}{m_{\rm up}}
\newcommand{\mi}{m_{\rm i}}
\newcommand{\mij}{m_{{\rm i,}j}}
\newcommand{\mpr}{m}
\newcommand{\np}{N_{\rm p}}
\newcommand{\mvsq}{\langle v^2\rangle}

\newcommand{\mvsqpm}{\langle v^2_{{\rm p}}\rangle}
\newcommand{\mvsqpl}{\langle v^2_{{\rm p}}\rangle_L}

\newcommand{\etar}{\eta_r}
\newcommand{\etav}{\eta_v}
\newcommand{\rhm}{r_{{\rm h}}}
\newcommand{\rhpm}{r_{{\rm hp}}}
\newcommand{\rhpl}{r_{{\rm hp},L}}

\newcommand{\rc}{r_{\rm 0}}

\newcommand{\rv}{r_{\rm v}}

\newcommand{\trh}{\tau_{\rm rh}}
\newcommand{\vrmsj}{v_{j0}^{\rm rms}}

\documentclass[useAMS,usenatbib, times]{mn2e}
\usepackage{amssymb,latexsym,graphicx,natbib,eufrak,times,amsmath}
\usepackage{hhline}

\title[The mass-to-light ratio of globular clusters]
  {Biases in the inferred mass-to-light ratio of globular clusters: no
  need for variations in the stellar mass function}
\author[Rosemary L. Shanahan \& Mark Gieles]
  {Rosemary L. Shanahan$^1$ \& Mark Gieles$^2$\\
$^1$ Institute for Astronomy, University of Edinburgh, Royal
Observatory, Blackford Hill, Edinburgh EH9 3HJ, UK.\\
$^2$ Department of Physics, University of Surrey, Guildford, GU2 7XH,
UK.
\\ 
}
\date{Accepted 2014 December 22.  Received 2014 December 20; in original form 2014 November 19}

\def\LaTeX{L\kern-.36em\raise.3ex\hbox{a}\kern-.15em
    T\kern-.1667em\lower.7ex\hbox{E}\kern-.125emX}

\begin{document}         
\maketitle
\begin{abstract}
From a study of the integrated light properties of 200 globular
clusters (GCs) in M31, Strader et al. found that the
mass-to-light ratios are lower than what is expected from simple
stellar population (SSP) models with a {`canonical' stellar
initial mass function (IMF)}, with the discrepancy being larger at
high metallicities.  We use dynamical multi-mass models, that include
a prescription for equipartition, to quantify the bias in the inferred
dynamical mass as the result of the assumption that light follows
mass. For a universal IMF and a metallicity dependent {\it present
day} mass function we find that the inferred mass from integrated
light properties systematically under estimates the true mass, and
that the bias is more important at high metallicities, as was found
for the M31 GCs.  { We show that mass segregation and a flattening
of the mass function have opposing effects of similar magnitude on the
mass inferred from integrated properties. This makes the mass-to-light
ratio as derived from integrated properties an inadequate probe of the
low-mass end of the stellar mass function.} There is, therefore, no
need for variations in the IMF, nor the need to invoke depletion of
low-mass stars, to explain the observations. Finally, we find that the
retention fraction of stellar-mass black holes (BHs) is an equally
important parameter in understanding the mass segregation bias. We
speculatively put forward to idea that kinematical data of GCs can in
fact be used to constrain the total mass in stellar-mass BHs in GCs.
\end{abstract}

\section{Introduction}
The stellar initial mass function (IMF) plays a vital role in
astrophysics and it is an ongoing debate whether the IMF is universal
for all metallicities \citep{2010ARA&A..48..339B,
2010Natur.468..940V,2013ApJ...771...29G}. Globular clusters (GCs) are
important test-beds for studies of the IMF and the present day mass
function (MF) because they are, in most cases, single aged stellar
populations with a single $\feh$, and the kinematics of the stars
within them is due to the gravity of the stars only. { Because most
of the GC mass is in low-mass stars, whereas the luminosity is due to
the high-mass stars, the GC mass-to-light ratio $\moverl \equiv M/\lv$
is consider to be a good probe of the shape of the stellar MF.}  Here $\lv$
is the $V$-band luminosity and $M$ is independently found from a
measurement of the cluster's radius and velocity dispersion.  Recent
observational advances have allowed us to measure the structural
parameters and kinematical properties of GCs in external galaxies up
to distances of several tens of Mpc \citep{2001MNRAS.326.1027S,
2006A&A...448..881B, 2006ARA&A..44..193B}, spanning a large range of
environments, thereby { potentially increasing the applicability of
$\moverl$} as a probe of the IMF.

{ For Galactic GCs, a slight positive correlation between $\moverl$
and $M$ was found by early work by \citet{1991A&A...252...94M}
and \citet{1993ASPC...50..357P}. A more significant correlation was
found from a larger sample of GCs by \citet{2014arXiv1411.1763K}.}
\citet{2005ApJS..161..304M} found an average $\langle\moverl\rangle\simeq
1.45$ for Milky Way GCs, which is slightly lower than the SSP
prediction of $\moverl\simeq2$, appropriate for low
metallicities. This small difference { and the trend with $M$} has
been { attributed to} the preferential escape of low-mass stars
over the tidal boundary, reducing $M$, but hardly affecting
$L_V$ \citep[e.g.][]{2008A&A...486L..21K}.

\citet[][hereafter S11]{2011AJ....142....8S} published the $\moverl$
values of 200 GCs in M31, spanning a range of metallicity of
$-2\lesssim \feh\ \lesssim 0$. They found that at low metallicities
the average $\moverl$ agrees well with the values from simple stellar
population (SSP) models, assuming a  Kroupa IMF, albeit
with a scatter of a factor of two. For increasing metallicities, a
growing difference between the dynamical $\moverl$ and the SSP models
was found, with $\moverl$ falling below the SSP values, by up to a
factor of 3 at $\feh\simeq0$. { Similar results were recently
obtained for Milky Way GCs \citep{2014arXiv1411.1763K}}. These results
are hard to explain by the preferential escape of low-mass stars,
because it implies that almost all metal-rich clusters have lost a
significant fraction ($\sim60\%-70\%$) of their initial mass.

An alternative interpretation is that the IMF gets more bottom light
(fewer low-mass stars) with increasing $\feh$. Before accepting the
possibility of a varying IMF, we need to establish that there are no
metallicity dependent biases as the result of dynamical evolution of
GCs in the $\moverl$ measurements.  A possible bias is the effect of
mass segregation on the measurements of the GC radius and velocity
dispersion. Most GCs are older than their half-mass relaxation
time-scale $\trh$ \citep[e.g.][]{1961AnAp...24..369H,
2011MNRAS.413.2509G}.  This means that they had time to evolve towards
equipartition and, as a result, the most massive objects (stellar
remnants and turn-off stars) are more centrally concentrated and move
with lower velocities.  For resolved GCs this effect can be taken into
account by fitting multi-mass
models \citep[e.g.][]{2010AJ....139..476P, 2012ApJ...755..156S}, which
are distribution function based models that include an approximate
prescription for equipartition \citep[][hereafter
GG79]{1976ApJ...206..128D, 1979AJ.....84..752G}. However, in
extragalactic samples where the individual stars are not resolved,
this is harder to do (although not impossible), and it is therefore
often assumed that light follows mass.

For a universal IMF, the stellar mass function at an age of 12\,Gyr
{\it does} depend on the metallicity of the stars {because of
differences in stellar evolution}, and as a result we may expect that
clusters with different metallicities, have different observable
properties.  The idea that a metallicity dependent bias in the
observationally derived  cluster properties may exist is not
new. \citet{2004ApJ...613L.117J} considered multi-mass King models,
with mass bins appropriate for different metallicities.  He showed
that as a result of mass segregation, clusters with the same half-mass
radius $\rhm$ and the same age, have smaller {\it half-light} radii in
projection, $\rhpl$, at higher metallicities. This is because at high
$\feh$ the turn-off mass is higher and as a result these stars and the
bright evolved stars, which have similar masses, are more centrally
concentrated than in metal-poor clusters.  Jord\'{a}n proposed that
this effect is responsible for the observed size difference between
metal-rich and metal-poor GCs \citep[e.g.][]{2001AJ....121.2974L}. An
apparent size difference between metal-rich and metal-poor GCs was
also found in $N$-body simulations \citep{2012MNRAS.427..167S} and
Monte Carlo simulations \citep{2012MNRAS.425.2234D} of clusters with
different metallicities.
 
In this Letter we use multi-mass King models to quantify the mass
segregation bias in $\moverl$ as a function of $\feh$ and for
different assumptions for the retention of remnants.
In \S~\ref{sec:models} we model  stellar
mass functions for different $\feh$ and describe the dynamical
models. In \S~\ref{sec:results} we present the results of a comparison
to the data by S11 and our conclusions and discussion are presented
in \S~\ref{sec:conclusions}.

\section{Description of the models}
\label{sec:models}
\subsection{Stellar mass functions at an age of 12 Gyr}
\label{ssec:mf}
Here we describe how we compute the stellar mass components for
clusters with different metallicities.  We create $\np=10^4$ star
particles with initial masses $\mi$ equally spaced in $\log\mi$
between a lower mass of $\mlo=0.2\,\msun$ and an upper mass of $\mup=
100\,\msun$.  We consider a power-law {(`canonical')} IMF
\begin{equation}
\frac{\dr N}{\dr\mi}=A\mi^{-\alpha},
\end{equation}
where $\alpha = 2.35$ and $A$ is a constant that normalises the total
number of stars in the cluster: $A=(1-\alpha)/(\mup^{1-\alpha}-
\mlo^{1-\alpha})$, for $\alpha\ne1$. Every star particle is given a
weight $w = A\mi^{1-\alpha}(\ln\mup-\ln\mlo)/\np$, such that
$\sum_{j=1}^{\np}w_j = 1$ and
$\sum_{j=1}^{\np}w_j{\mij}\simeq0.68\,\msun$ is the mean mass of the
IMF, which is similar to the value for more realistic IMFs that roll
over at low masses to flatter slopes and are truncated at
$\mlo=0.1\,\msun$. {For our purpose this suffices, because the
behaviour of the multi-mass models is sensitive only to the total mass
in low-mass stars, and insensitive to the mean mass of the low-mass
stars (see Section~5 of GG79). Tests confirm that our results are insensitive to
this choice.}  {To study the effect of mass
segregation in a system with a bottom light IMF, or a cluster that is
depleted in low-mass stars because of dynamical evolution, we also
consider a double power-law (`flat') IMF
\begin{equation}
\frac{\dr N}{\dr\mi}=
\begin{cases}
B, \hspace{1.1cm}\mlo<\mi\le1\,\msun,\\
A\mi^{-\alpha}, \hspace{0.5cm}1\,\msun<\mi<\mup,
\end{cases}
\end{equation}
where $B=A$ to make the function continuous. We normalise the
high-mass end in the same way as for the `canonical' IMF (i.e. the
same $A$), so the reduction in mass due to flattening can be found
straightforwardly.  }

We evolve every star particle to an age of 12\,Gyr using  SSE 
\citep{2000MNRAS.315..543H}  for 11 metallicities, linearly spaced in $-2 \le \feh \le
0$. We split the data into bins corresponding to the stellar types at
12\,Gyr: main sequence stars (MS), evolved stars (EV), white dwarfs
(WD), neutron stars (NS) and black holes (BH).  The logarithmic
spacing\footnote{Approximating the mean mass in a bin by the central
logarithmic value is only exact for a power-law with index
$-1.5$. From a convergence test we find that for $10^4$ stars
our model properties such as the total $M$ of the MF are accurate to
about $10^{-5}$.} in $\mi$ and the fast {\small{SSE}} code allow us to
efficiently model MFs of old stellar populations that are well sampled
at all evolutionary stages and hence present day masses $\mpr$.

\begin{figure}
\centering
\includegraphics[width=8cm]{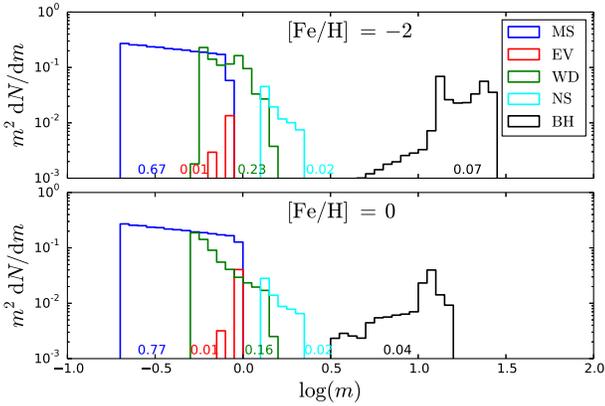}
\caption{Mass functions for two of our evolved stellar
  populations: one metal poor (top panel), one metal rich (bottom
  panel). Each distribution is split into the five stellar types: main
  sequence stars (MS), evolved stars (EV), white dwarfs (WD), neutron
  stars (NS), and black holes (BH). The numbers are the fraction of
  the total mass in each component.}
\label{fig:mass_functions}
\end{figure}

The resulting MFs for a metal-rich and a metal-poor cluster {with
a `canonical' IMF} are shown in Fig.~\ref{fig:mass_functions}. From
this figure we see that at $\feh=-2$ the turn-off mass is slightly
lower ($0.8\,\msun$) than at $\feh=0$ ($1\,\msun$) and that the
average mass of WDs is higher at low $\feh$ ($0.77\,\msun$) than at
high $\feh$ ($0.65\,\msun$). Also, the contribution of remnants to the
total mass is higher at low $\feh$.  Although these differences are
small, they are important for the distribution of the visible stars,
as we will show in the next section.

For our dynamical models we create 13 mass components: we split the MS
stars into 5 bins that are equally spaced in $\log(\mpr$), the WDs
into 5 bins that are equally spaced in $m$, the EV stars, NSs and,
finally, the BHs.  With the $\feh$ dependent mass functions in
place, we can now describe the dynamical multi-mass models.

\subsection{Multi-mass models}
In this section we describe the dynamical models that we use to
quantify the effect of mass segregation on the inferred dynamical mass
from observations.  For a self-gravitating system in virial
equilibrium, the mass $M$ can be written as a function of the
mass-weighted mean-square velocity of the stars, $\mvsq$, and the
virial radius of the cluster, $\rv$, as $M = 2\mvsq\rv/G$. Here $\rv$
is defined as $\rv\equiv -GM^2/(2U)$, where $G$ is the gravitational
constant and $U$ is the total potential energy of the cluster.  Our
focus is on extra-galactic cluster samples for which individual
stars are not resolved, so we write the virial relations in terms of
the {\it luminosity-weighted} mean-square velocity in projection
$\mvsqpl$ and the radius containing half the total {\it luminosity} in
projection $\rhpl$ (or, `effective radius')

\begin{equation}
M =  2\frac{\eta}{\etar\etav}\frac{\mvsqpl\rhpl}{G}.
\label{eq:mdyn}
\end{equation}
Here we introduce three factors $\eta$, $\etar$ and $\etav$ to convert
3-dimensional properties to observable quantities, namely
\begin{align}
\displaystyle\eta = \frac{\mvsq}{\mvsqpm}\frac{\rhm}{\rhpm}\frac{\rv}{\rhm},
\hspace{0.5cm}\etar =  \frac{\rhpl}{\rhpm},
\hspace{0.5cm}\etav = \frac{\mvsqpl}{\mvsqpm}.\label{eq:eta}
\end{align}
Here $\mvsqpm$ is the {\it mass-weighted} mean-square velocity in
projection and $\rhpm$ is the radius containing half the {\it mass} in
projection.  The value of $\eta$ depends on the model, and for a
\citet{1911MNRAS..71..460P} model with an isotropic velocity
distribution  $\eta\simeq3\times(4/3)\times(5/4)=5$, and this
value is relatively insensitive to the choice of the model
\citep{1987degc.book.....S, 2010ARA&A..48..431P}. From now on we 
assume that uncertainties in $\eta$ are small and can be ignored.

The factors $\etar$ and $\etav$ equal unity if light follows mass,
i.e. if $\moverl$ is the same everywhere in the cluster. If this
assumption is made for clusters for which this does not hold, then the
product $\etar\etav$ is the error that is made in the mass
estimate. \citet{2007MNRAS.379...93H} finds from $N$-body simulations
that $\eta_r\simeq0.5$ which implies an error in $M$ of a factor of
two, confirming that mass segregation is an important effect to
consider.

We derive the bias factor $\etar\etav$ from isotropic multi-mass King
models as described in detail in GG79. The distribution function of
each component $m_j$ (section~\ref{ssec:mf}) is
\begin{equation}
  f_j(E) = A_j\left[\exp\left(-\frac{E}{\sigma_j^2}\right) -1\right].
\end{equation} 
Here $E=\frac{1}{2}v^2+\phi$, and $\phi$ is the (lowered) specific
 potential and $\sigma_j$ is a scale velocity\footnote{We note that
 only in the limit of infinite concentration $\sigma_j$ equals the
 central 1D velocity dispersion $\vrmsj$ of bin $j$. For realistic concentration values the ratio $\vrmsj/\sigma_j<1$ and is mass dependent. Multi-models models are therefore
 {\it not} in equipartition. In forthcoming studies (\citealt{sollima2015}; Peuten et al., in
 prep) we show that multi-mass models describe the phase space
 distribution of stars in $N$-body systems very well.  } that is
 chosen to depend on $m_j$ as $\sigma_j \propto m_j^{-1/2}$.

{From the \citet{2005MNRAS.362..799M} SSP models we find that
about 65\% of the $V$-band light comes from EV stars and 35\% from the
MS stars. For a typical mass-luminosity relation for MS stars we
estimate that about 70\% of the MS luminosity is emitted by stars in
the mass range of EV stars ($\gtrsim0.65\,\msun$). We therefore assume
that the dynamical properties of EV stars are the ones that one infers
from integrated light studies.}

\begin{figure}
\centering \includegraphics[width=8cm]{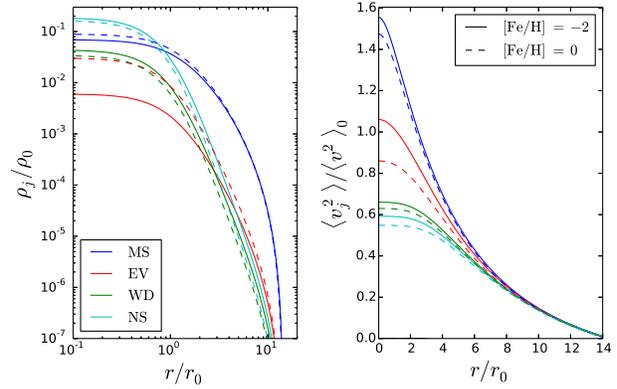} 
  \caption{Density and velocity dispersion profiles for two of our GC
           models with $W_0\!=\!7$. Although our multi-mass models
           have 13 bins, for clarity we have only plotted the middle
           mass bin values for MS and WD. Solid lines:
           $\feh=-2$. Dashed lines: $\feh=0$. }
\label{fig:density_velocity_profiles}
\end{figure}

In Fig.~\ref{fig:density_velocity_profiles} we show the resulting
density and velocity profiles for a metal-rich and a metal-poor
cluster without BHs.  For both clusters the EV stars are more
centrally concentrated and move with lower velocities than the MS
stars which implies that both $\etar<1$ and $\etav<1$.  The
underestimation of $M$ is more severe for the metal-rich cluster than
for the metal-poor cluster.  This is because at high $\feh$, the
turn-off mass is higher, and the remnants are less massive (see
Fig.~\ref{fig:mass_functions}), allowing the bright stars to segregate
further to the centre of the cluster \citep[see
also][]{2004ApJ...613L.117J,2012MNRAS.427..167S}.

We now need to make a choice for the dimensionless central potential
$W_0$. We choose $W_0$ such that the EV stars have a ratio of the core
radius $\rc$ (or King radius) over $\rhm$ that is equal to what is
found for the M31 GCs by S11.  We obtained the values for $\rc/\rhpl$
of the GCs in M31 from S11 and found that there is no correlation
between $\rc/\rhpl$ and $\feh$. We therefore adopt 3 values: the mean
observed value and the values corresponding to the mean and the plus
and minus $1\sigma$ standard deviation. {Because $\rc$ for the
individual components has a different meaning then $\rc$ in single
mass models because, we decide to define $\rc$ as the distance to the
centre where the density of EV stars has dropped to 1/3 of the central
density, as is the case in single mass models with
$W_0\gtrsim5$. Finally, we use $\rhpl=0.75\rhm^{\rm EV}$, where $\rhm^{\rm EV}$ is the (3D) half-mass radius of the EV stars.}

Using these three $\rc/\rhpl$ values, we produce ten models with $W_0$
ranging from 6 to 24, for each $\feh$. We then interpolate to obtain
the values of $W_0$ to obtain the desired $\rc/\rhpl$ value.  A sample
of the $W_0$ values used can be found in Table 1.

\begin{table}
\centering
        \caption{A sample of the $W_0$ values used for different
        $\feh$ and the different assumptions for  retention
        of NSs and BHs {for a `canonical' IMF}.}  

        \begin{tabular}{|c|c||c|c|c|} \hline [Fe/H]
        & $\rc/\rhpl$ &All BH \& NS&No BH &No BH \& NS\\
            \hline
            \hline
                   & 0.45 & 15.9 & 8.86 & 8.41  \\
            $-2$   & 0.29 & 17.5 & 11.1 & 10.3\\
                 & 0.13 & 19.8& 14.2& 12.9\\
             \hline             
                & 0.45 &23.0 & 10.8 & 10.5  \\
            0  & 0.29 & 24.4 & 13.4 & 12.4 \\
                & 0.13 & 25.8 & 16.6 & 15.1 \\
             \hline
        \end{tabular}
\end{table}

\section{Results}
\label{sec:results}

\begin{figure*}
        \centering \includegraphics[width=16cm]{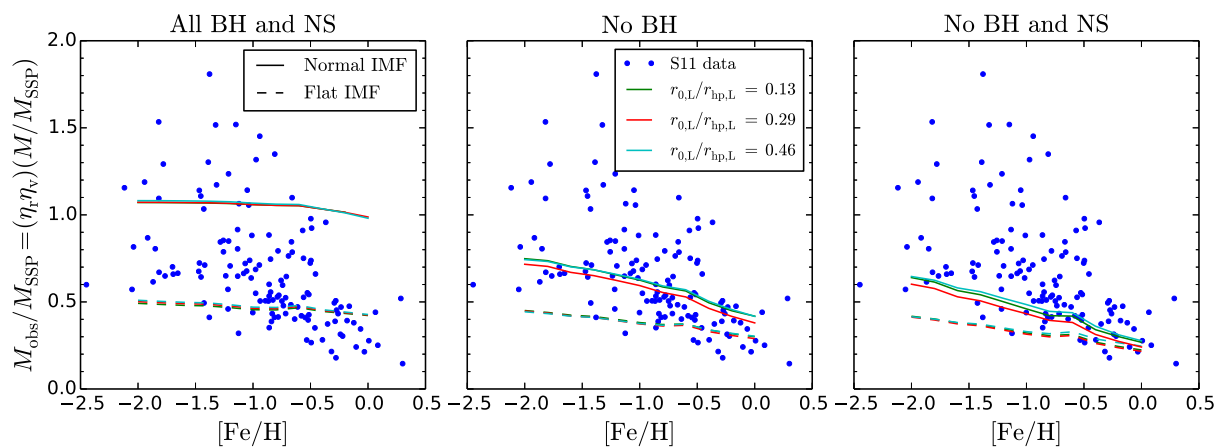}
        \caption{Ratio of inferred mass $\mobs$ over SSP mass $\mssp$versus $\feh$ {for a `canonical' IMF
                 (solid lines) and flat IMF (dashed lines)}. The blue
                 points are the corresponding values for the M31 GCs
                 taken from S11. The three lines in each of the plots
                 show the results for models designed to `look' as
                 similar to the S11 GCs as possible, using the mean
                 $\rc/\rhpl$ value and two $\rm\sigma$ values. For the
                 `canonical' IMF $M/\mssp=1$, and for the `flat' IMF
                 $M/\mssp \simeq 0.41$}.  \label{fig:mass_metal}
\end{figure*}

For each $\feh$, $\rc/\rhpl$ and different remnant fractions,
we calculate the ratio of the mass as it would be derived from the
evolved stars, $\mobs$, over the true mass $M$, i.e. $\etar \etav$
(equation~\ref{eq:eta}). In this calculation, we used exclusively the
contribution from the evolved stars to derive $\mobs$, in order to be
consistent with the observations. {For the models with a
`canonical' IMF $\mobs/M = \mobs/\mssp$, and for the models with a
`flat' IMF we multiplied $\mobs/M$ by $M/\mssp\simeq0.41$, slightly dependent on $\feh$, to obtain
$\mobs/\mssp$.}  The results are shown in
Fig.~\ref{fig:mass_metal}. The three panels show the results of three
assumptions for the remnant retention fraction, from left to right:
all BHs and all NSs retained, no BHs retained and all NSs retained, no BHs
and no NSs retained. The results for a model in which all BHs are
retained and no NSs are retained were found to be almost identical to
the model in which all NSs and all BHs were retained.

We divided the inferred $\moverl$ of each GC in M31 by its
corresponding value of the SSP model (Fig.~1 of S11), based on a
Kroupa IMF, to obtain $\mobs/\mssp$ for M31 GCs, {where $\mssp$ is
the present day mass of a stellar population with a `canonical' IMF}.
{We note that there is a slight inconsistency in the comparison
between our models and the data because the $\moverl$ values of the SSP models 
are based on a single value for the retention fraction of NSs and BHs. However,
as we show in Fig.~\ref{fig:mass_functions} the total mass in NSs and
BHs never exceeds 9\% of the present day $M$, so this effect is small.}

{If all NSs and BHs are retained (left panel), there is hardly any
bias because of mass segregation and the correct $M$ is inferred from
the EV stars. Without NSs and BHs (right panel),
$0.25\lesssim\mobs/\mssp\lesssim0.5$ for both MFs. The mass
segregation term ($\etar\etav$) is higher when the mass function is
flatter, while $M/\mssp$ is smaller. The two effects appear to roughly
cancel, such that the resulting ratio $\mobs/\mssp$ is similar. This
means that a low $\mobs/\mssp$ can not be interpreted as a flattening
of the MF, nor the IMF, because a mass segregated model with a
`canonical' IMF gives similar results.}

The two extremes of the retention fraction models (i.e. all the BH and
NS retained, and none of the BH and NS retained) encompass the
majority of the GC values, and it is possible to interpret the spread
in the data as a spread in BH retention.

The $\feh$ dependence in the middle and right panels can be explained
with some simple stellar evolution arguments. From
Fig.~\ref{fig:mass_functions} we see that at lower metallicities the
turn-off mass is lower and there is more mass in remnants, and the
average mass of white dwarf mass and black holes is higher \citep[see
also the discussion in ][]{2012MNRAS.427..167S}.  As a result, in a
low-metallicity cluster, the remnants are more efficient in pushing
the evolved stars out, which increases their velocities and their
half-light radius, such that $\etav$ and $\etar$ are closer to unity.

These results suggest that the discrepancy between the observed GC
masses of S11 and the SSP models, can be explained by metallicity
dependent mass segregation effects and a spread in the retention
fraction of NSs and BHs. {Our results suggest that the low-mass
end of the MF can not be constrained from integrated light
properties.}

\section{Conclusions and discussion}
\label{sec:conclusions}
In this study we show that the discrepancy between the observed
mass-to-light ratio $\moverl$ and simple stellar population (SSP)
models as found in a sample of 200 globular clusters (GCs) in M31 by
S11, can be explained by mass segregation.  We modelled GCs with a
range of metallicities and with different remnant retention
fractions. Our key results are:

\begin{itemize}
\item For clusters without stellar-mass black holes (BHs) and neutron stars, the
  dynamical mass derived from evolved stars ($\mobs$) of a mass
  segregated cluster underestimates the true mass $M$, and this bias
  is stronger at high $\feh$ {(factor of $\sim4$)} than at low
  $\feh$ {(factor of $\sim2$)}.  This is due to the higher
  turn-off mass and the lower white dwarf masses at high $\feh$ which
  cause the evolved stars to be more centrally concentrated and move
  with lower velocities, compared to turn-off stars in low $\feh$
  clusters with the same mass density profile \citep[see
  also][]{2004ApJ...613L.117J,2012MNRAS.427..167S}.

\item The presence of BHs has a large effect on the derived $\mobs$,
  in that its value is closer to the real $M$ for higher retention
  fractions. Despite the fact that a large fraction of BHs probably
  gets ejected from GCs as the result of supernova kicks, it is
  important to consider the BHs in dynamical mass modelling of GCs. A
  small remaining BH population can survive for as long as 10
  half-mass relaxation times \citep[$\trh$,
  ][]{2013MNRAS.432.2779B}. The recent discovery of two BH candidates
  in M22 \citep{2012Natur.490...71S} confirms the need to consider BHs
  when deriving $\moverl$.

\item {For clusters  with a `flat' mass function, we find that 
  the effect of mass segregation on $\mobs/M$ is less important, such
  that the ratio  $\mobs/\mssp$, with $\mssp$ is the present day mass for a canonical IMF, is similar to $M/\mssp$ models with a
  `canonical' IMF. This makes $\moverl$ insensitive to the low-mass
  end of the MF. }
\end{itemize}

S11 find that the discrepancy between $\moverl$ and the SSP models is
more important for clusters with lower mass and for clusters with
shorter $\trh$, which was also found for GCs in the Milky
Way \citep{1991A&A...252...94M, 2014arXiv1411.1763K} and Centaurus
A \citep[NGC 5128,][]{2007A&A...469..147R}. This is consistent with
the scenario put forward in this Letter, because mass segregation is
more important for clusters with shorter $\trh$, i.e. those with lower
mass.

\citet{2014arXiv1409.3235Z} report a bi-modality in $\moverl$, after
all clusters have been aged to a common age of 10\,Gyr
($\moverlten$). They determine dynamical masses from integrated light
properties of a sample of star clusters with different ages and
metallicities and in different galaxies.  The authors interpret their
result as evidence for two distinct IMFs. Although we do not discuss
the mass segregation bias in their sample in detail here, we caution
to invoke the need of a varying IMF in a study based on integrated
light measurement, before biases as a result of mass segregation have
been fully quantified.

The reduced $\moverl$ of MW GCs \citep{2005ApJS..161..304M} and other
galaxies has been explained by the depletion of low-mass stars as the
result of dynamical ejections \citep{2008A&A...486L..21K}. This effect
is also more important for clusters with shorter $\trh$/lower mass,
therefore degenerate with the mass segregation bias reported here.  We
note that preferential depletion of low-mass stars requires mass
segregation, but not vice versa and it is therefore necessary to
consider the biases due to mass segregation {together with the
effect of the depletion of low-mass stars. Here we showed that the two
effects have opposing effects on $\mobs/\mssp$, such that $\moverl$ is
 insensitive to the slope of the MF.}

The $\moverl$ values of young (few 100\,Myrs) massive
($\gtrsim10^6\,\msun$)  clusters in merger remnants are
consistent with the predictions of SSP models with a `canonical' IMF
\citep{2006A&A...448..881B}. Although these values were derived from
integrated light properties, these clusters are significantly younger
than their two-body relaxation time scale, such that mass segregation
is likely not important yet.

Finally, the strong dependence of $\mobs/M$ on the retention fraction
of stellar mass BH opens the exciting opportunity to use kinematic
data of GCs to determine the total mass in stellar mass BHs, thereby
placing complementary constraints on BH retention in {Milky Way
GCs}.

\section*{Acknowledgements}
This work was carried out as part of the ISIMA programme on
Gravitational Dynamics held in July 2014 at CITA, Toronto. We
acknowledge Pascale Garaud for the organisation, the financial support
and for providing a stimulating environment. MG thanks the Royal
Society for financial support in the form of a University Research
Fellowship (URF) and the European Research Council (ERC-StG-335936,
CLUSTERS). RLS acknowledges the support of the Science and Technology
Facilities Council (STFC) via the award of an STFC Studentship. RLS
thanks Anna Lisa Varri for her support and guidance throughout this
work. We thank Miklos Peuten, Antonio Sollima, Anna Lisa Varri and
Alice Zocchi for discussions during the `Gaia Challenge' meeting
and {Diederik Kruijssen,Vincent H\'{e}nault-Brunet, Jay Strader
and Nate Bastian for comments on the manuscript.} { Finally, we
thank the referee for constructive comments that helped to improve the
paper.}

\bibliographystyle{mn2e}

\end{document}